\documentclass[%
 reprint,
 amsmath,amssymb,
 aps,
floatfix,
]{revtex4-2}

\usepackage{amsmath}
\usepackage{graphicx} 
\usepackage{bm}
\usepackage{xcolor}
\usepackage{placeins} 
\usepackage{ulem}
\newcommand{\mm}{\textrm{mm}}
\newcommand{\s}{\textrm{s}}

\begin{document}

\title{Identifying non-equilibrium fluctuations in Intracellular Motion Using Recurrent Neural Networks}

\author{Tomas Basile}
\author{Natascha Leijnse}
\author{Malte Slot Lauridsen}
\author{Younes Farhangi Barooji}
\author{Amin Doostmohammadi}
\email{doostmohammadi@nbi.ku.dk}
\author{Karel Proesmans}
\email{karel.proesmans@nbi.ku.dk}
 
\affiliation{%
 Niels Bohr Institute, University of Copenhagen, Copenhagen, Denmark.
}%
\date{\today}

\begin{abstract}

Distinguishing active from passive dynamics is a fundamental challenge in understanding the motion of living cells and other active matter systems. Here, we introduce a framework that combines physical modeling, analytical theory, and machine learning to identify and characterize active fluctuations from trajectory data. We train a long short-term memory (LSTM) neural network on synthetic trajectories generated from well-defined stochastic models of active particles, enabling it to classify motion as passive or active and to infer the underlying active process. Applied to experimental trajectories of a tracer in the cytoplasm of a living cell, the method robustly identifies actively driven motion and selects an Ornstein–Uhlenbeck active noise model as the best description. Crucially, the classifier's performance on simulated data approaches the theoretical optimum that we derive, and it also yields accurate estimates of the active diffusion coefficient. This integrated approach opens a powerful route to quantify non-equilibrium fluctuations in complex biological systems from limited data. 
\end{abstract}
\maketitle

Living systems operate fundamentally out of equilibrium, continuously consuming and dissipating energy to sustain essential biological functions. This persistent non-equilibrium state is crucial for life at the cellular level~\cite{non_eq_indispensable}, driving key processes such as DNA replication~\cite{noneq_dna,proofreading}, intracellular transport~\cite{Intracellular_transport}, and cellular organization~\cite{cell_mechanics,equilibrium_cytoskeletal}.

A prominent class of non-equilibrium systems is active matter~\cite{active_matter_2,active_matter_4,Irreversibility_in_active_matter}, which consists of particles that extract energy from their surroundings and convert it into motion. Biological examples include cytoskeletal networks~\cite{active_cyto}, nuclear fluctuations~\cite{Nucleus-fluctuations}, swarming bacteria~\cite{bacteria}, multicellular layers~\cite{andersen2025evidence}, and, on larger scales, flocking birds and human crowds~\cite{bird_flocks}. A key challenge in studying biological motion is determining whether an observed trajectory results from active dynamics or if it can be explained by passive thermal fluctuations. Experimentally distinguishing between these two cases is often difficult, as active fluctuations can be subtle or masked by environmental noise. 
It generally is even more challenging to determine which type of active noise model is best at describing the system.

In this article, we tackle this problem using a machine learning-based method to classify active noise in experimental trajectories. We train a type of recurrent neural networks, known as long short-term memory (LSTM) neural network on synthetic data generated from models of passive and active motion, including the Active Ornstein-Uhlenbeck Particle (AOUP)~\cite{AOUP,Irreversibility_in_active_matter,Irreversibility_in_active_matter_2} and the Rotational Brownian Particle (RBP)~\cite{active_brownian,active_brownian2}. By applying this model to experimental data of a tracer particle inside the cytoplasm of a cell, we identify the presence of active noise in living cell trajectories and show that it is best described by an AOUP model, thereby demonstrating the method’s effectiveness in real-world applications.

This approach builds on previous studies that have applied machine learning to non-equilibrium physics, such as using neural networks to infer entropy production~\cite{Learning-EP,tur1,learning_non_eq} or detect the arrow of time~\cite{ML_arrow}. Unlike those works, our focus is on identifying active noise in biological motion. By providing a flexible classification tool, our method opens new possibilities for analyzing a wide range of experimental datasets in biological systems.

To model the motion of an active particle, we consider a colloidal particle moving in one dimension while suspended in an aqueous solution at thermal equilibrium with temperature 
$T$. The particle is subject to a deterministic conservative force $f(x)=-\partial_xU(x)$, derived from a potential 
$U(x)$, as well as random fluctuations arising from both thermal and active sources. The latter represents  energy-consuming processes that drive the system out of equilibrium, distinguishing active motion from passive Brownian motion.

Neglecting inertial effects (i.e., assuming an overdamped regime), the particle's dynamics are described by the Langevin equation~\cite{Irreversibility_in_active_matter}:
\begin{equation}
\dot{x} = \dfrac{1}{\gamma} f(x) + \sqrt{2D} \xi(t) + \sqrt{2D_a} \eta(t),
\label{eq: x_active_noise}
\end{equation}
where $\gamma$ is the hydrodynamic friction coefficient of the particle, $D$ is the thermal diffusion coefficient, $\xi(t)$ is the thermal white noise, satisfying $\langle \xi(t)\rangle=0$, $\langle \xi(t)\xi(t')\rangle=\delta(t-t')$  and $D_a$ is the active diffusion coefficient. The term $\eta(t)$ models the active fluctuations, which can take different forms depending on the underlying active process. 

A commonly used and biologically relevant model for $\eta(t)$ is the Ornstein-Uhlenbeck process~\cite{AOUP,Irreversibility_in_active_matter,Irreversibility_in_active_matter_2}:
\begin{equation}
\dot{\eta}(t) = - \dfrac{1}{\tau_a} \eta(t) + \dfrac{1}{\tau_a} \zeta(t),
\label{eq:AOUP_eta}
\end{equation}
where $\zeta(t)$ is a Gaussian white noise with zero mean and delta-correlated fluctuations, that is, $\langle \zeta(t)\rangle =0$ and $\langle \zeta(t) \zeta(t') \rangle = \delta(t-t')$.
When active noise is modeled this way, the full system given by Eqs.~\eqref{eq: x_active_noise} and ~\eqref{eq:AOUP_eta} is known as an Active Ornstein-Uhlenbeck Particle (AOUP). This model captures the persistence in motion over characteristic timescales set by $\tau_a$, one of the key features of active matter systems~\cite{peyret2019sustained}. 

Alternatively, active noise can also be modeled by rotational Brownian particles (RBP)~\cite{active_brownian, active_brownian2,Active_Brownian_Analytical}. Here, the active noise is defined as $\eta(t) = \cos (\theta(t))$, with $\theta(t)$ evolving as a simple Brownian motion:
\begin{equation}
\dot{\theta}(t) = \sqrt{2D_{rot}}\; \zeta(t),
\end{equation}
 where $D_{rot}$ is a constant that determines the speed at which the particle changes direction.
This formulation represents active fluctuations with persistent rotational diffusion, a feature commonly associated with self-propelled particles such as bacteria~\cite{active_brownian}, and active colloids~\cite{BechingerRevMod}. 

\begin{figure}
\includegraphics[width=8cm]{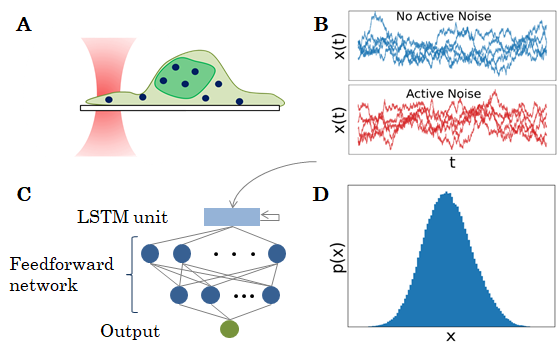}
\caption{ \label{fig1} (A) Schematic of the experimental setup. (B) Example of simulated trajectories of a particle evolving according to the AOUP model, shown on the top for the passive case (that is, with $D_a = 0$) and on the bottom for the active case. (C) Schematic of the machine learning classification workflow: a given trajectory is first processed by an LSTM unit, followed by two fully connected layers of a feedforward network, which outputs a single value between $0$ and $1$, representing the predicted probability that the trajectory has active noise. (D) Position histogram for the cellular trajectory studied in this paper. }
\end{figure}

{\it LSTM detects activity from limited data.} Having established a theoretical framework for modeling active noise, we now apply our method to real trajectory data of a tracer particle inside the cytoplasm of a living cell, that is controlled by an optical tweezer. Our goal is to determine whether the motion of a tracer particle inside the cytoplasm of a cell exhibits active fluctuations and, if so, whether it can be well described by either the AOUP model or the RBP model. For that, we will study the real trajectory data in order to generate similar artificial data and use that to train an LSTM to classify active noise. 

We begin by analyzing the one-dimensional trajectory of the trapped tracer particle. As seen in Fig.~\ref{fig1}, a histogram of the experimentally observed position of the particle reveals an almost perfectly Gaussian distribution, suggesting that the cell experiences a harmonic potential, of the form  $U(x) =kx^2/2$, with a corresponding deterministic force $f(x) = -kx$. 

To get a rough idea of the numerical values of parameters associated with the dynamics, we will compare the correlation function associated with the experimental data with that of AOUP and RBP in a harmonic potential under steady-state conditions. In the supplemental material A, we show that for AOUP:
\begin{align}
\nonumber\langle x(t) x(0) \rangle & = \left( \dfrac{D}{k_{\gamma}} + \dfrac{D_a}{k_{\gamma}(1-k_{\gamma}^2 \tau_a^2)} \right) e^{-k_{\gamma}|t|}\\
& \quad\quad - \dfrac{D_a \tau_a}{1-k_{\gamma}^2 \tau_a^2}  e^{-|t|/\tau_a}, \label{eq:corr}
\end{align}
\noindent where $k_{\gamma} = k/\gamma$.
Fitting this expression to the experimental correlation function (see Fig.~\ref{fig:correlation} in the supplemental material) allows us to estimate the key parameters of the AOUP model for the cell's motion, yielding $k_\gamma = 157 \pm 7\s^{-1}$, $\tau_a = 0.33 \pm 0.011\s$, $D = (7 \pm 1.2 ) \times 10^{-4} \;  \mm^2\s^{-1}$, and $D_a = (6.3  \pm 0.5) \times 10^{-3} \;
 \mm^2\s^{-1}$.  These values provide a physically motivated parameter space for generating synthetic trajectories to train our machine learning model.

To construct the training dataset, we generate trajectories with random values of $k_\gamma$, $\tau_a$, and $D$, sampled uniformly within ranges determined from the fitted experimental parameters: $k_{\gamma} \in [50\s^{-1},250\s^{-1}], \tau_a \in [0.05\s,0.8\s]$, and $D \in [10^{-6}\mm^2\s^{-1},0.002\mm^2\s^{-1}]$. Half of the trajectories are generated with $D_a = 0$ (passive motion), while the other half include nonzero active diffusion with $D_a \in [10^{-6}\mm^2\s^{-1}, 0.01\mm^2\s^{-1}]$.

\begin{figure}
\includegraphics[width=8cm]{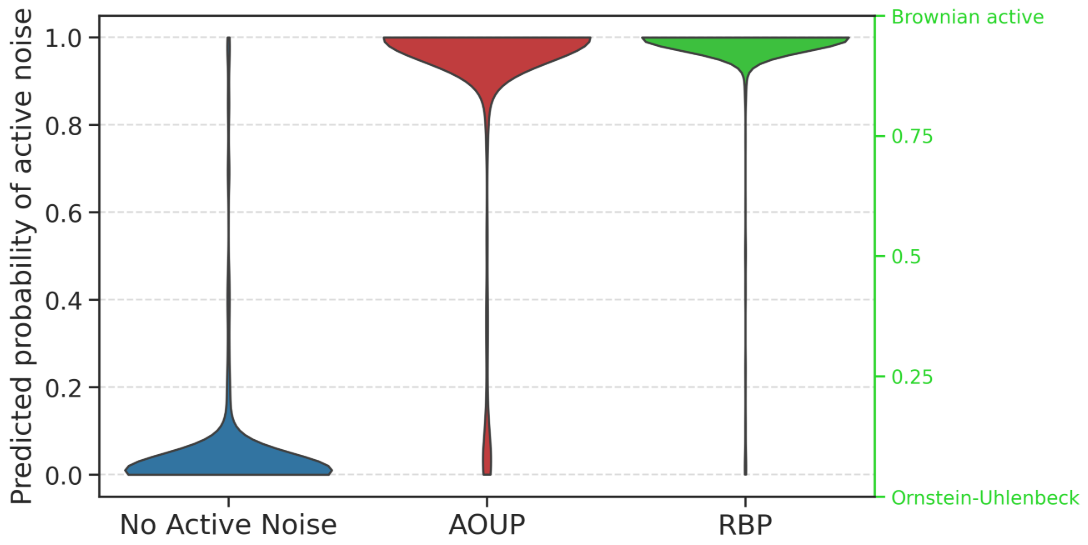}
\caption{The LSTM classifier was trained on simulated data to distinguish between passive and active trajectories based on their time series. The model assigns a low probability of having active noise to passive trajectories (blue, "No Active Noise") and a high probability to trajectories generated with Ornstein-Uhlenbeck active noise (red, "AOUP"). The classifier successfully differentiates passive from active cases, with minimal overlap between distributions. The rightmost plot (green, "RBP") shows the results of applying a separate LSTM trained to distinguish between rotational brownian noise (labeled with a 1) and Ornstein-Uhlenbeck noise (labeled with a 0) to simulated data with rotational brownian noise. The result shows that it correctly distinguishes the different active noises and classifies the rotational brownian trajectories with values close to 1.}
\label{fig2}
\end{figure}

After training the LSTM model to classify these trajectories as active or passive, we apply it on test trajectories created in the same way. We find that it performs well in distinguishing between active and passive motion. The results are presented in Fig.~\ref{fig2}, showing violin plots of the predicted probabilities of having active noise for trajectories created without active noise (left) and with active noise following the AOUP model (center). The results show that the network correctly gives low probabilities to trajectories without active noise (resulting in an average probability of $0.046$) and high probabilities to trajectories with active noise (resulting in an average of $0.986$). 

{\it Cellular trajectories exhibit non-equilibrium (active) noise.} Having trained the LSTM on artificial trajectories, we apply it to the real cellular data, yielding a result of $0.997$, which shows with a very high level of confidence that the dynamics of the tracer particle inside the cytoplasm of a cell exhibits active noise. These results also confirm that the method can reliably detect active noise in trajectory data, providing a robust tool for identifying non-equilibrium dynamics in living cells and other active matter systems. 

\FloatBarrier

{\it Model Selection Favors AOUP over Rotational Brownian Noise.} To determine the type of active noise, we now do a similar analysis for RBP.
To construct the trajectories, we follow the same method as before to get ranges for the parameters $k,D,D_a,D_{rot}$. In this case, the analytical correlation function is~\cite{Active_Brownian_Analytical}:
\begin{align}
\nonumber\langle x(t) x(0) \rangle & = \left(\dfrac{D}{k_{\gamma}} + \dfrac{D_a D_{rot}}{k_{\gamma} (D_{rot}^2 - k^2)} \right) e^{-k_{\gamma}|t|} \\
& \quad\quad - \dfrac{D_a}{D_{rot}^2-k_{\gamma}^2}  e^{-D_{rot}|t|}.
\label{eq:corr_BAM}
\end{align}
As before, we fit this expression to the experimental result of the correlation function. Since the expression has the same shape as the one in Eq.~\eqref{eq:corr}, the fit is the same, but yielding now the following parameters $k/\gamma = 157 \pm 7 \s^{-1}$, $D_{rot} = 3.0 \pm 0.05\; \mm^2 \s^{-1}$, $D_{a} = 0.021 \pm 0.002 \; \mm^2 \s^{-1}$, and $D=(7\pm 1.2) \times 10^{-4} \;\mm^2 \s^{-1}$. 

We create the training set of trajectories with random values of $k, D_{rot},D,D_a$ taken respectively from $[50\s^{-1},250\s^{-1}], [1\s^{-1},10\s^{-1}], [10^{-6} \mm^{2}\s^{-1}, 0.002 \mm^2 \s^{-1}]$ and $[10^{-6} \mm^2 \s^{-1}, 0.1 \mm^2 \s^{-1}]$. Given these trajectories, we train the LSTM to distinguish between RBP and AOUP systems. Fig.~\ref{fig2} presents the result of applying this LSTM to test trajectories with rotational Brownian noise, showing that it correctly gives a confident prediction (an average of $0.985$) that the trajectories stem from an RBP system, instead of AOUP system.
When this LSTM is used on the real cellular data, the trajectory is classified with a confidence of $0.983$ as being best described by the AOUP model. Therefore, we conclude with a very high level of confidence that the dynamics of a tracer in the cytoplasm of a cell is best described by an AOUP.

{\it Classifier performance nears the theoretical optimum.} A natural question to ask at this point is how much room there is to improve the quality of the classification. To answer this question, we now test it thoroughly with simulated trajectories. We consider a particle under a harmonic potential $U(x) =k x^2/2$, and generate trajectories with parameters set at $D = 0.2 \mm^2 \s^{-1}, \tau_a = 0.5\s$ and $k_{\gamma} = 0.1\s^{-1}$, and a time step of $\Delta t = 0.01\s$. We simulate $5,000$ trajectories with $D_a = 0$ (no active noise) and $5,000$ with $D_a = 0.1\mm^2\s^{-1}$ (active noise present). 

\begin{figure}
\includegraphics[width=8cm]{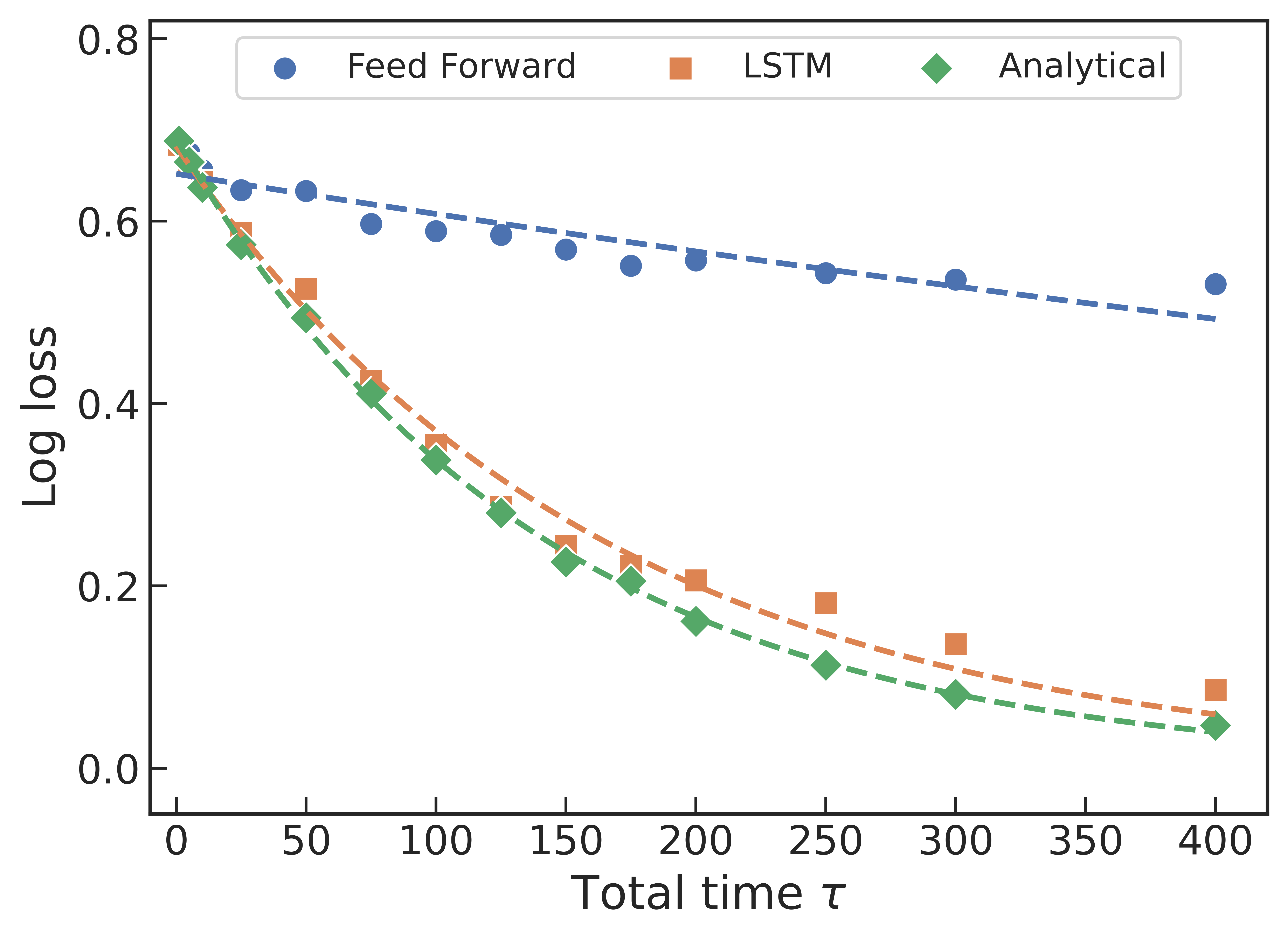}
\centering
\caption{Classifier performance nears the theoretical optimum. Log-loss of the binary classification of trajectories with and without Ornstein-Uhlenbeck active noise, as a function of total trajectory length. The LSTM and analytical model show strong improvement in classification accuracy as trajectory length increases, whereas the feedforward model exhibits only limited improvement.}
\label{fig3}
\end{figure}

Given this dataset, we train both a feedforward neural network and a LSTM neural network  to classify trajectories as having active noise (labeled with a $1$) or not  (labeled with a $0$). The neural networks output a probability between $0$ and $1$, representing their confidence that a given trajectory contains active noise.
To evaluate the performance, we test the models on
$2,000$ previously 
unseen trajectories (half with active noise and half without), and quantify it using the Log loss metric~\cite{NN_Book}, defined as
\begin{align}
\label{logloss}
\text{Log loss} = - \dfrac{1}{N} \sum_{k=1}^N  \left( y_k \log (p_k) + (1-y_k) \log (1-p_k) \right),
\end{align}
where $N$ is the number of test trajectories, $y_k$ is the true label for the k-th trajectory (i.e., $1$ if the particle is active and $0$ if it is passive). and $p_k$ the network's output. 
This whole process is then repeated for a variety of total durations, and the results as a function of trajectory duration are shown in Fig.~\ref{fig3}. 

Fig.~\ref{fig3} also presents the theoretically calculated best possible performance of such a classifier on the test trajectories. The details of this theoretical derivation are given in the supplemental material, section B. There we prove that if we have a set of trajectories, half without active noise and half consisting of AOUPs with constant $D_a$, then the most reasonable estimate of the probability that a trajectory  $\underline{x}$  has active noise is given by
\begin{widetext}
\begin{equation}
P(D_a = D_a' | \underline{x} ) = \dfrac{1}{1+   C \exp \bigg\{ - \dfrac{D_a}{4D^2} \displaystyle \int_0^{\tau} dt \int_0^{\tau}\;  dt' (\dot{x}_t - g_t) \Gamma_{\tau}(t,t') (\dot{x}_{t'}-g_{t'})  \;  \bigg\} },
\label{eq:final_theo}
\end{equation}
\end{widetext}
with $C$ given by
\begin{equation}
C = \dfrac{1}{2} L^{-1/4} e^{- \tau k_- / (2\tau_a)}  \sqrt{ 4 \sqrt{L} + (1-\sqrt{L})^2 (1-\rho^2)  },
\end{equation}
$g(x(t),t) =  f(x(t),t)/\gamma$, $L = 1+D_a/D$, $k_{\pm} = 1 \pm \sqrt{L}$, $\rho = \exp\left({-\sqrt{L}\tau/\tau_a}\right)$, and $\Gamma_{\tau}(t,t')$ is defined in Eq.~(\ref{eq: gamma}) in supplemental material B. From this, one can determine the minimal log-loss associated with the dataset.

{\it Trajectory length drives exponential gains in accuracy.} Fig.~\ref{fig3} highlights an important trend: as the total trajectory length increases and more information is present for the classifiers, the efficiency of both the LSTM and the analytical classifier improves exponentially, while the feedforward neural network shows very limited improvement. This suggests that one can classify the noise arbitrarily well, provided that one has sufficiently long trajectories. Furthermore, the log-loss of the LSTM is very close to that of the analytical calculation, showing that our network classifies the data in an almost optimal way.

So far, we have successfully determined whether a system is out of equilibrium and identified the type of active noise. However, a key quantity for further analysis in non-equilibrium systems is the active diffusion coefficient, $D_a$, which can also be estimated using an LSTM.

\begin{figure}
\includegraphics[width=8cm]{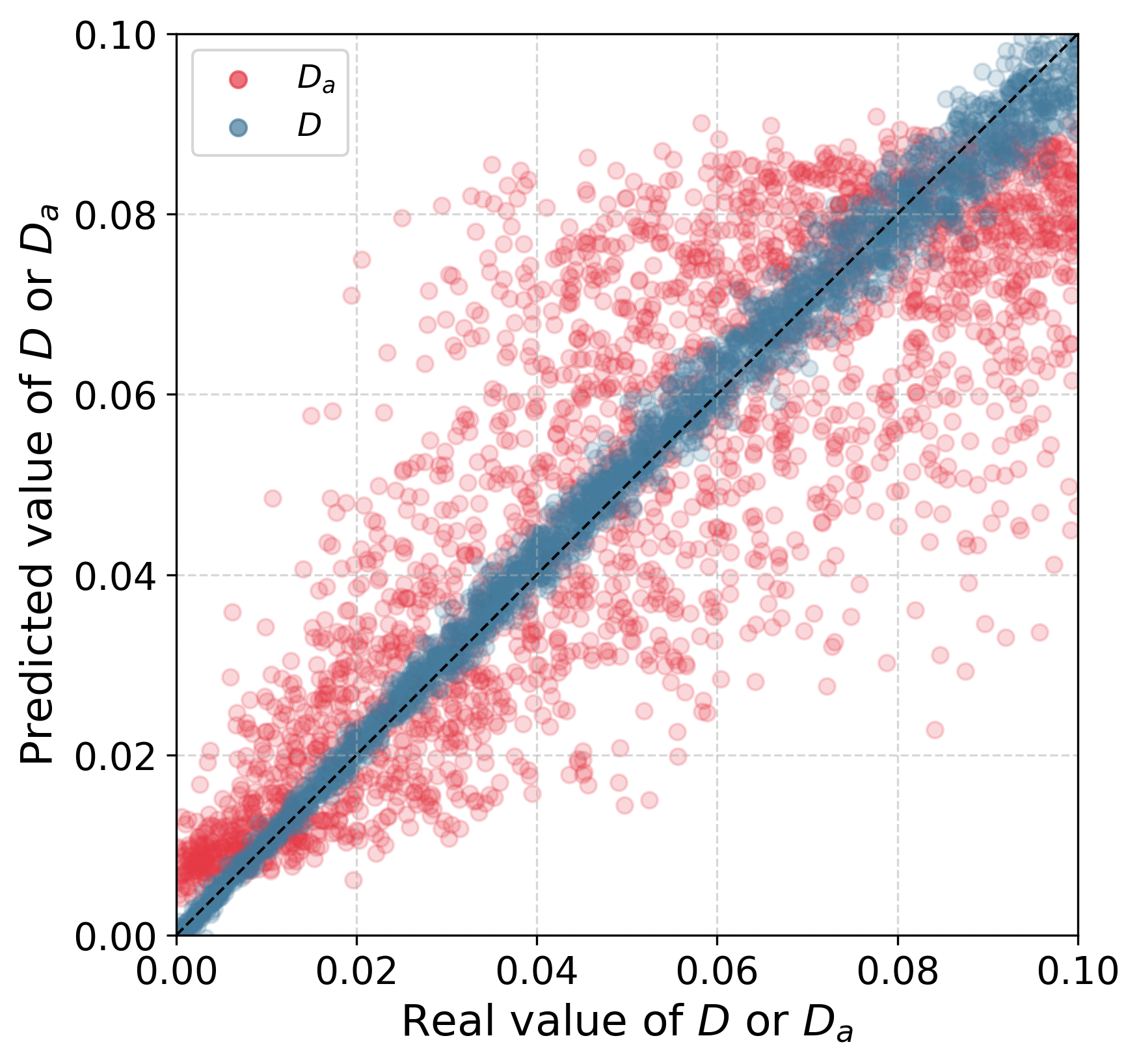}
\centering
\caption{Results of LSTMs in predicting $D$ and $D_a$ on the test set, compared to their real values. }
\label{fig4}
\end{figure}

{\it Estimating the diffusion coefficients from single trajectories.} To investigate this, we generated $10,000$ synthetic trajectories with varying values of $D$ and $D_a$ sampled separately from the range $[10^{-4}\mm^2\s^{-1}, 0.1\mm^2\s^{-1}]$. We then trained two separate LSTMs: one to predict $D$ and another to predict $D_a$
from a given trajectory. The results on the test set, shown in Fig.~\ref{fig4}, demonstrate that while the model 
is reasonably good at estimating $D_a$, it is much better for $D$. Specifically, the normalized root mean squared error (calculated as the root of the mean square difference between predictions and true values, divided by the average of the true values) obtained for $D_a$ is $0.27$, while for $D$ it is $0.05$.
This discrepancy is to be expected, as 
$D$ can in principle be extracted from a single trajectory with an arbitrarily level of precision using $D \simeq {\langle \Delta x^2\rangle}/{2\Delta t}$, for an arbitrary short time-window $\Delta t$. In contrast, different values of 
$D_a$ can lead to very similar trajectories, leading to an inherent uncertainty on its estimation..
Nevertheless, when applied to the experimental cell trajectory, the network gives results of $D = 5.9 \times 10^{-4}\mm^2\s^{-1}$ and $D_a = 4.3 \times 10^{-3} \mm^2 \s^{-1}$, 
which are considerably close to the ones obtained before by fitting the correlation function.

We introduced a framework that unites stochastic theory, numerical modeling, recurrent neural networks, and experimental data to identify and characterize active fluctuations in living cells. After training a Long Short-Term Memory (LSTM) network on synthetic trajectories generated from physically interpretable models, we successfully classified the presence of active noise in real intracellular motion. Importantly, this classification is grounded in theory, allowing us to go beyond detection and infer the specific nature of the underlying active process. By benchmarking against analytical predictions, we further showed that the accuracy of our approach approaches the theoretical optimum, demonstrating that recurrent neural networks can extract essentially all information available in finite trajectories. Applied to tracer motion in the cytoplasm, the method robustly revealed non-equilibrium fluctuations and identified the active Ornstein–Uhlenbeck process as the best description of the dynamics. More broadly, tailoring the training data to alternative models enables systematic extension to other cells and active matter systems, providing a general route to connect noisy experimental trajectories with the mechanisms that generate them.

\begin{acknowledgments}
It is a pleasure to acknowledge valuable feedback on the manuscript from Etienne Fodor. A. D. acknowledges funding from the Novo Nordisk Foundation (grant No. NNF18SA0035142 and NERD grant No. NNF21OC0068687), Villum Fonden (Grant no. 29476), and the European Union (ERC, PhysCoMeT, 101041418). Views and opinions expressed are however those of the authors only and do not necessarily reflect those of the European Union or the European Research Council. Neither the European Union nor the granting authority can be held responsible for them.
\end{acknowledgments}

\bibliography{references} 

\clearpage
\begin{widetext}
\section{End Matter}

\subsection{A. Correlation function of an AOUP in a harmonic potential}
\label{smA}

In this section we calculate the correlation $\langle x(t) x(t') \rangle$ for an AOUP on a harmonic potential, which is described by
\begin{align*}
\dot{x}(t) = -k_{\gamma}x(t) + \sqrt{2 D_a} \eta(t) + \sqrt{2D} \xi(t), \\
\dot{\eta}(t) = - \dfrac{1}{\tau_a} \eta(t) + \dfrac{1}{\tau_a} \zeta(t),
\end{align*}
where $k_{\gamma} = k/\gamma$, $\xi$ is an unbiased Gaussian white-noise, such that $\langle \xi(t) \rangle = 0$ and  $\langle \xi(t) \xi(t') \rangle = \delta(t-t')$, and similarly for $\zeta(t)$.

\subsubsection*{Correlation for $\eta(t)$}
Before calculating $\langle x(t)x(t') \rangle$, we need to calculate the correlation for $\eta(t)$, that is $\langle \eta(t) \eta(t') \rangle$. To do so, we start by formally solving the differential equation for $\eta(t)$, giving
\begin{equation}
\eta(t) =  \eta(0) e^{-t/\tau_a} + \dfrac{1}{\tau_a}
\int_0^{t} \zeta(s) e^{(s-t)/\tau_a} ds.
\label{eq:sol_eta}
\end{equation}
Now we can easily get the ensemble mean of $\eta(t)$ by using that $\langle \zeta(t) \rangle = 0$:
\begin{align}
\langle \eta(t) \rangle = \langle \eta(0) \rangle e^{-t/\tau_a}.
\label{eq:eta_mean}
\end{align}
Therefore, we see that after a long time, the variable $\eta(t)$ will have a mean of $0$, that is, once the particle has reached steady state, we can assume that $\langle \eta(t) \rangle = 0$. 
Now we can calculate the correlation $\langle \eta(t) \eta(t') \rangle$. To do so, we use the result of Eq.~\eqref{eq:sol_eta} to get:
\begin{align*}
\eta(t) \eta(t') & =  \eta(0)^2 e^{-(t+t')/\tau_a}  + \eta(0) \dfrac{1}{\tau_a} \int_0^{t'} \zeta(s) e^{(s-t'-t)/\tau_a} ds + \eta(0) \dfrac{1}{\tau_a} \int_0^{t} \zeta(s) e^{(s-t-t')/\tau_a} ds \\
& \;\;\;\; + \dfrac{1}{\tau_a^2} \int_0^t \int_0^{t'} \zeta(s) \zeta(s') e^{(s+s'-t-t' )/\tau_a} ds' ds.
\end{align*}
Then, we take the expected value of this quantity and use that $\langle \zeta(s) \rangle = 0$, $\langle \zeta(s) \zeta(s') \rangle = \delta(s-s')$. We also use that $\langle \eta(0) \zeta(s) \rangle = \langle \eta(0) \rangle \langle \zeta(s) \rangle = 0$ because $\eta$ at time $0$ is uncorrelated to the value of $\zeta$ at later times. Therefore:
\begin{align}
\nonumber \langle \eta(t) \eta(t') \rangle &=  e^{-(t+t')/\tau_a } \langle \eta(0)^2 \rangle + \dfrac{1}{\tau_a^2} \int_0^t\int_0^{t} \delta(s-s') e^{(s+s'-t-t' )/\tau_a} ds' ds \\
\nonumber & = e^{-(t+t')/\tau_a } \langle \eta(0)^2 \rangle + \dfrac{1}{\tau_a^2} \int_0^te^{(2s-t-t' )/\tau_a}  ds \\
& = e^{-(t+t')/\tau_a} \left( \langle \eta(0)^2 \rangle - \dfrac{1}{2\tau_a} \right) + \dfrac{1}{2\tau_a} e^{-|t'-t|/\tau_a}.
\label{eq:corr_eta_res}
\end{align}
We can also directly obtain the steady state second moment $\langle \eta(t)^2 \rangle \bigg|_{t \rightarrow \infty}$ by setting $t' = t$ and then letting $t \rightarrow \infty$, giving as a result $
\langle \eta(t)^2 \rangle = \dfrac{1}{2\tau_a}$.
Therefore, if the system begins in steady state, the term $\langle \eta(0)^2 \rangle$ in Eq.~\eqref{eq:corr_eta_res} is equal to $\dfrac{1}{2\tau_a}$ and so the expression for the correlation in steady state reduces to:
\begin{align}
\langle \eta(t) \eta(t') \rangle = \dfrac{1}{2 \tau_a} e^{- |t'-t|/\tau_a}.
\label{eq:corr_eta}
\end{align}

\subsubsection{Correlation for $x(t)$}

Now we calculate the correlation $\langle x(t) x(t') \rangle$. To do so, we first solve formally the equation for $x(t)$, giving:
\begin{align}
x(t) =   x(0) e^{-k_{\gamma}t} + \sqrt{2D_a} e^{-k_{\gamma}t} \int_0^t \eta(s) e^{k_{\gamma}s} ds + \sqrt{2D} e^{-k_{\gamma}t} \int_0^t \xi(s) e^{k_{\gamma}s}ds.
\label{eq:Formal_solution}
\end{align}
Now we can compute the product of $x(t)x(t')$, which is:
\begin{align*}
x(t)  x(t') &= x(0)^2 e^{-k_{\gamma}(t+t')} \; +\;  x(0) \sqrt{2D_a} \; e^{-k_{\gamma}(t+t')}  \int_0^t\eta(s) e^{k_{\gamma}s} ds\;  + \; x(0) \sqrt{2D} e^{-k_{\gamma}(t+t')} \int_0^t \xi(s) e^{k_{\gamma}s} ds \\
& \;\;\; +\; x(0) \sqrt{2D_a} e^{-k_{\gamma}(t+t')} \int_0^{t'} \eta(s') e^{k_{\gamma}s'}ds' \; + \; 2D_a e^{-k_{\gamma}(t+t')} \int_0^t \int_0^{t'} \eta(s) \eta(s') e^{k_{\gamma}(s+s')} ds'ds \\
& \;\;\; +\; 2 \sqrt{D D_a} e^{-k_{\gamma}(t+t')} \int_0^t \int_0^{t'} \xi(s) \eta(s') e^{k_{\gamma}(s+s')} ds'ds \\
& \;\;\; + \; x(0) \sqrt{2D} e^{-k_{\gamma}(t+t')} \int_0^{t'} \xi(s') e^{k_{\gamma}s'} ds' \;+\; 2 \sqrt{DD_a} e^{-k_{\gamma}(t+t')} \int_0^t \int_0^{t'} \xi(s') \eta(s) e^{k_{\gamma}(s+s')} ds'ds \\
& \;\;\; + \; 2D e^{-k_{\gamma}(t+t')} \int_0^t \int_0^{t'} \xi(s) \xi(s') e^{k_{\gamma}(s+s')} ds'ds.
\end{align*}
Then, we need to take the mean of this, for which we use that $\langle \eta(s) \rangle = 0$ in steady state, $\langle \xi(s) \rangle = 0$ and that $\xi$ and $\eta$ are uncorrelated, so that $\langle \xi(s') \eta(s) \rangle = \langle \xi(s') \rangle \langle \eta(s)\rangle = 0$. 
We also use that $x(0)$ and noises at later times are uncorrelated, so that $\langle x(0) \xi(s) \rangle = \langle x(0) \rangle \langle \xi(s) \rangle = 0$.
With this, we can see that the second,  third, fourth, sixth, seventh, and eight terms in the result for $x(t)x(t')$ are equal to $0$ and therefore we get:
\begin{align}
\nonumber \langle x(t) x(t') \rangle &=  \langle x(0)^2 \rangle e^{-k_{\gamma}(t+t')} \;+\; 2D_a e^{-k_{\gamma}(t+t')} \int_0^t \int_0^{t'}  
\langle \eta(s) \eta(s') \rangle
e^{k_{\gamma}(s+s')} ds' ds + 2D e^{-k_{\gamma}(t+t')} \int_0^t \int_0^{t'} \langle \xi(s) \xi(s') \rangle e^{k_{\gamma}(s+s')} ds'ds \\
& = \langle x(0)^2 \rangle e^{-k_{\gamma}(t+t')}  + \dfrac{D_a}{\tau_a}e^{-k_{\gamma}(t+t')}  \int_0^t \int_0^{t'}  e^{k_{\gamma}(s+s')} e^{-|s'-s|/\tau_a} ds'ds + 2D e^{-k_{\gamma}(t+t')} \int_0^t \int_0^{t'} \delta(s-s') e^{k_{\gamma}(s+s')} ds'ds,
\label{eq:resxtxtp}
\end{align}
where we used the result obtained for $\langle \eta(s) \eta(s') \rangle$.
The second integral can be calculated straightforwardly, giving $\dfrac{1}{2k_{\gamma}} (e^{2k_{\gamma}t}-1)$. On the other hand, the first integral is more complicated, requiring first to separate the interval $[0,t']$ into $[0,t]$ and $[t,t']$.
\begin{align*}
\int_0^t \int_0^{t'}  e^{k_{\gamma}(s+s')} e^{-|s'-s|/\tau_a} ds'ds &= \int_0^t \left[ \int_0^t e^{k_{\gamma}(s+s')} e^{-|s'-s|/\tau_a} ds' +  e^{\Lambda_+ s}  \int_t^{t'} e^{\Lambda_- s'} ds' \right]ds \\
& = \dfrac{1}{\Lambda_+} \left[ \dfrac{1}{2k_{\gamma}} \left( e^{2k_{\gamma}t} -1 \right) - \dfrac{1}{\Lambda_-} \left(e^{\Lambda_- t} - 1 \right) \right] + \dfrac{1}{\Lambda_-} \left[ \dfrac{e^{\Lambda_- t}}{\Lambda_+} \left( e^{\Lambda_+ t} - 1 \right)  - \dfrac{1}{2k_{\gamma}} \left( e^{2k_{\gamma}t} -1 \right)  \right]     \\
 & \;\;\;\; \quad\quad+\dfrac{1}{\Lambda_+ \Lambda_-}  \left( e^{\Lambda_+ t} - 1 \right)\left( e^{\Lambda_- t'} - e^{\Lambda_- t} \right).
\end{align*}
where $\Lambda_- := k_{\gamma}-1/\tau_a$ and $\Lambda_+ := k_{\gamma}+1/\tau_a$. Now that we have the results for both integrals in Eq.~\eqref{eq:resxtxtp}, we can substitute them to get the correlation. Then, after some manipulation, we get:
\begin{align}
\nonumber\langle x(t) & x(t') \rangle  =  \left( \langle x(0)^2 \rangle + \dfrac{D-Dk_{\gamma}\tau_a + D_a}{k_{\gamma}(-1+k_{\gamma}\tau_a)} \right) e^{-k_{\gamma}(t+t')} +  \left( \dfrac{D}{k_{\gamma}} + \dfrac{D_a}{k_{\gamma}(1-k_{\gamma}^2 \tau_a^2)} \right) e^{-k_{\gamma}(t'-t)} + \dfrac{D_a \tau_a}{1-k_{\gamma}^2 \tau_a^2} \; e^{-t/\tau_a - k_{\gamma}t'}  \\
& \;\;\; + \dfrac{D_a \tau_a}{1-k_{\gamma}^2\tau_a^2} \; e^{-t'/\tau_a - k_{\gamma}t} - \dfrac{D_a \tau_a}{1-k_{\gamma}^2 \tau_a^2} \; e^{-1/\tau_a (t'-t)}.
\label{eq:corr_casi_final}
\end{align}
However, we are only interested in the result after a long time has passed, so that we have reached steady state. To get this, we first need to see what is the value of $\langle x(0)^2 \rangle$ in steady state.
To do so, we set $t = t'$ and then $t \rightarrow \infty$, getting as a result
\begin{align*}
\langle x(t)^2 \rangle = \dfrac{D + Dk_{\gamma}\tau_a + D_a}{k_{\gamma}(1+k_{\gamma}\tau_a)}.
\end{align*}
Therefore, if we start from steady state, we can set this as the value for $\langle x^2(0)\rangle$. Substituting this into Eq.~\eqref{eq:corr_casi_final}, and disregarding the terms that become $0$ if $t$ and $t'$ are large enough, we get the steady state correlation:
\begin{align}
\langle x(t) x(t') \rangle = \left( \dfrac{D}{k_{\gamma}} + \dfrac{D_a}{k_{\gamma}(1-k_{\gamma}^2 \tau_a^2)} \right) e^{-k_{\gamma}(t'-t)} - \dfrac{D_a \tau_a}{1-k_{\gamma}^2 \tau_a^2} \; e^{-1/\tau_a (t'-t)} \\
\label{eq:corr_x_final}
\end{align}

\subsubsection{Correlation Fit}
\begin{figure}[h!]
\includegraphics[width=8cm]{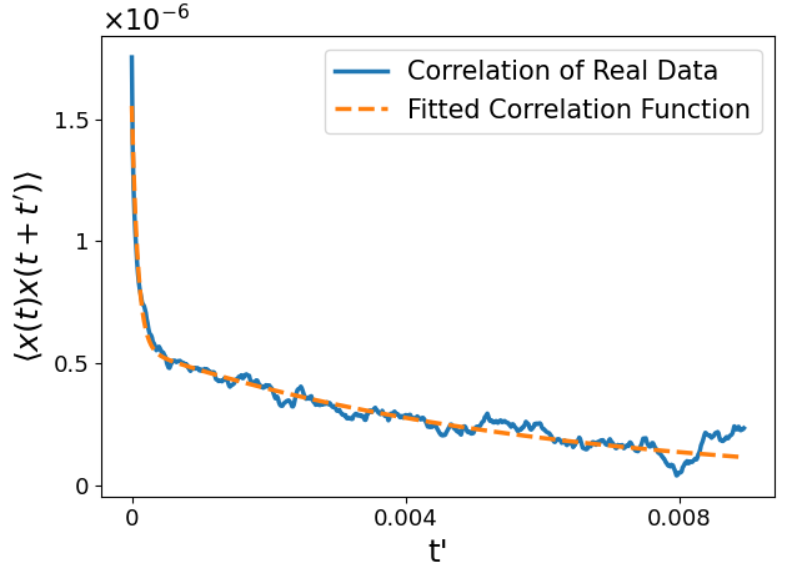}
\caption{ \label{fig:correlation}  Result of calculating the correlation $\langle x(t) x(t+t')\rangle$ as a function of $t'$ for the experimental cell trajectory (in blue). Also shown is the result of fitting Eq.~\eqref{eq:corr} (or Eq.~\eqref{eq:corr_BAM}, since it has the same shape) to the data (in orange).}
\end{figure}

\subsection{B. Theoretical calculation of $P(D_a = D_a' | \underline{x})$}
\label{smb}

In this section, we outline the theoretical calculation for performing the classification  of active noise from a trajectory. As before, we consider two types of trajectories, both characterized by the same parameters $k/\gamma$, $D$, and $\tau_a$. However, one category corresponds to trajectories without active noise ($D_a = 0$), while the other category has active noise with a fixed active diffusion coefficient $D_a'$. Given a trajectory, our goal is to calculate the probability that it belongs to the category with active noise, $P(D_a = D_a' \mid \underline{x} )$, which is the same quantity we attempted to determine using neural networks. As we will explain later, $\underline{x}$ represents the trajectory omitting the first point $x_0$. 

Using Bayes' theorem, we can express this probability as:
\begin{align*}
P(D_a = D_a' \; | \underline{x}) &= \dfrac{P( \underline{x}|D_a= D_a') P(D_a = D_a')}{P( \underline{x})} \\
& =\dfrac{P( \underline{x}|D_a= D_a') P(D_a = D_a')}{P( \underline{x}|D_a= D_a') P(D_a = D_a') + P( \underline{x}|D_a= 0) P(D_a = 0)}  \\
& = \dfrac{P( \underline{x}|D_a = D_a')}{P( \underline{x} | D_a = D_a') + P( \underline{x}|D_a = 0)},
\end{align*}
where in the second line we used the law of total probability. We further assume that, a priori, we don't have any preference about the trajectory, so that $P(D_a = 0) = P(D_a = D_a') = 1/2$.  Then, after simplifying we get:
\begin{align}
P(D_a = D_a' |  \underline{x}) = \dfrac{1}{1 + \dfrac{P( \underline{x} | D_a = 0)}{P( \underline{x}|D_a  =D_a')}}.
\label{eq:bayes_result}
\end{align}
Therefore, finding the quantity that we care about, the probability that a given trajectory has active noise, can be done by calculating both $P( \underline{x}|D_a = 0)$ and $P( \underline{x}|D_a = D_a')$ and then using Eq.~\eqref{eq:bayes_result}. 

$P( \underline{x}|D_a = 0)$ is the probability that the given trajectory $ \underline{x}$ is observed for a model without active noise, while $P( \underline{x}|D_a = D_a')$ is the probability that the trajectory is observed for a model with active noise. Both of these probabilities require path integrals to be calculated.

To do said calculations, we start by stating the Onsager-Machlup integral~\cite{machlup,Intro-to-stochastic}, which gives the probability density of a trajectory for a stochastic process. Then, we apply it to the specific system of an active Ornstein-Uhlenbeck particle, following a similar route as~\cite{Irreversibility_in_active_matter}.

\subsubsection*{Onsager-Machlup Integral}

We start by considering that our system is described by a stochastic differential equation, such as:
\begin{align*}
\dot{x}(t) = a(x(t),t) + b \xi(t),
\end{align*}
with $b$ a constant and $\xi(t)$ a Gaussian noise satisfying $\langle \xi(t) \xi(t') \rangle = \delta(t-t')$.

The particle follows a stochastic trajectory $x(t)$, during the time interval $t\in[0,\tau]$, and we assume that the initial position of the particle has some fixed value $x(0) = x_0$. We discretize time into $N$ subintervals of length $\Delta t$, so that the discretized trajectory is $\overline{x} := \{ x_0, x_1, x_2, \cdots, x_N\}$. We also define the part of the trajectory after the initial position as $\underline{x} := \{x_1, x_2, \cdots, x_N \}$.

The quantity we wish to calculate is the probability of having a certain trajectory $\underline{x}$ given the initial position $x_0$, that is $P(\underline{x}|x_0)$. This probability can be found to be~\cite{Intro-to-stochastic}:
\begin{align}
\label{eq:machlup}
P(\underline{x} | x_0) d\underline{x} = C_0 \exp\left( - \int_0^{\tau} dt  \left[ \dfrac{[(\dot{x}(t_n) - a(x_n,t_n))]^2}{2 b^2} + \dfrac{1}{2}  \dfrac{\partial}{\partial x_{n}'}a(x_n', t_n)   \right] \right),
\end{align}
where $C_0 = \lim_{\Delta t \rightarrow 0}  \left( \dfrac{1}{b\sqrt{2\pi \Delta t}} \right)^N$.

\subsubsection*{Calculation of $P(\underline{x}|x_0)$ without active noise}

Here we use the Onsager-Machlup integral~\cite{Intro-to-stochastic} to calculate the probability of observing a given trajectory for an overdamped system without active noise. In said case, the Langevin equation is:
\begin{align*}
\dot{x} = g(x(t),t) + \sqrt{2D} \xi(t),
\end{align*}
with $g(x(t),t) = \dfrac{1}{\gamma} f(x(t),t)$ and $\xi(t)$ an unbiased Gaussian white noise such that $\langle \xi(t) \xi(t') \rangle = \delta(t-t')$. 
Then, according to the Onsager-Machlup integral in Eq.~\eqref{eq:machlup} (with $a \rightarrow g$ and $b \rightarrow \sqrt{2D}$), we have that:
\begin{align*}
P(\underline{x} \;|\; x_0) = C_0 \exp \bigg\{ - \int_0^{\tau} dt \left[ \dfrac{(\dot{x}_t - g_t)^2}{4D} + \dfrac{1}{2} \dfrac{\partial g_t}{\partial x} \right]  \bigg\},
\end{align*}
where the subindex $t$ means that the variable is evaluated at time $t$,
and the normalization constant $C_0$ is given by
\begin{align*}
C_0 = \lim_{\Delta t \rightarrow 0} \left( \dfrac{1}{\sqrt{4  \pi D \Delta t }} \right)^N. 
\end{align*}

\subsubsection*{Calculation of $P(\underline{x}|x_0)$ with active noise}

Now we consider the case of an active Ornstein-Uhlenbeck particle, which is governed by the following equations: 
\begin{align*}
\dot{x}(t) &= g(x(t),t) + \sqrt{2D_a} \eta(t) + \sqrt{2D} \xi(t) \\
\dot{\eta}(t) &= - \dfrac{1}{\tau_a} \eta(t) + \dfrac{1}{\tau_a} \zeta(t),
\end{align*}
where $g(x(t),t) := \dfrac{1}{\gamma} f(x(t),t)$. 
In this case, since we have active noise, the trajectory of $x$ is non-Markovian and we can't apply directly the Onsager-Machlup integral. However, the evolution of the combined set of variables $(x,\eta)$ is Markovian and we can find a probability $p(\underline{x}, \underline{\eta} \; | \; x_0,\eta_0 )$. Then, we can integrate out the variable $\eta$ to get a probability density for $x(t)$. 

The Onsager-Machlup integral for the two variables $(x,\eta)$ is:
\begin{align}
P(\underline{x}, \underline{\eta} \; | \; x_0, \eta_0)  = C_1 \exp\left( - \int_0^{\tau} dt \left[ \dfrac{(\dot{x}_t - g_t - \sqrt{2D_a} \eta_t)^2}{4D} + \dfrac{(\tau_a \dot{\eta}_t + \eta_t)^2}{2} + \dfrac{1}{2} \dfrac{\partial g_t}{\partial x} \right] \right),
\label{eq:machlup-2}
\end{align}
and the normalization factor is
\begin{align*}
C_1 = \left( \dfrac{1}{\sqrt{2\pi \Delta t}} \right)^{2N} \left( \dfrac{\tau_a}{\sqrt{2D}} \right)^N e^{\tau/(2\tau_a)}.
\end{align*}

Now we need to integrate out the $\eta$.
From the law of total probability, we know that we can integrate it out by doing: 
\begin{align}
\label{eq:peta}
p(\underline{x} | x_0) = \int \mathcal{D} \underline{\eta}\; d \eta_0 \; p(\underline{x}, \underline{\eta} | x_0, \eta_0) p_0(\eta_0|x_0).
\end{align}
Here $p_0(\eta_0|x_0)$ is the initial distribution of $\eta_0$, which may depend on the value of $x_0$. However, we assume it to be independent from $x_0$, so that:
\begin{align}
p_0(\eta_0 | x_0) = p_s(\eta_0) = \sqrt{\dfrac{\tau_a}{\pi}} e^{-\tau_a \eta_0^2}.
\label{eq:Gaussian_eta}
\end{align}
The distribution for $\eta_0$ is a Gaussian with mean $0$ and standard deviation of $\dfrac{1}{\sqrt{2 \tau_a}}$. The assumption that $\eta_0$ is independent from $x_0$ can come from assuming that $t=0$ is the moment when the particle is placed into the medium, or that the correlations between $x$ and $\eta$ are not very significant. 

Substituting Eq.~\eqref{eq:Gaussian_eta} and Eq.~\eqref{eq:machlup-2} into Eq.~\eqref{eq:peta}, (and using the abbreviation $\overline{\eta} = \eta_0 \cup \underline{\eta}$), we get:
\begin{footnotesize}
\begin{align*}
& p(\underline{x}|x_0) = C_1 \sqrt{\dfrac{\tau_a}{\pi}} \int \mathcal{D} \overline{\eta} \;  e^{-\tau_a \eta_0^2}  \exp\left( - \int_0^{\tau} dt \left[ \dfrac{(\dot{x}_t - g_t - \sqrt{2D_a} \eta_t)^2}{4D} + \dfrac{(\tau_a \dot{\eta}_t + \eta_t)^2}{2} + \dfrac{1}{2} \dfrac{\partial g_t}{\partial x} \right] \right) \\
& = C_1' K \int \mathcal{D} \overline{\eta}\; e^{-\tau_a \eta_0^2} \exp\left( -\int_0^{\tau} dt \left[ \dfrac{-2 \sqrt{2D_a}(\dot{x}_t - g_t) \eta_t + 2D_a \eta_t^2}{4D} + \dfrac{\tau_a^2 \dot{\eta}_t^2 + 2\tau_a \dot{\eta}_t \eta_t +\eta_t^2}{2} \right] \right), 
\end{align*}
\end{footnotesize}
where $C_1' = \sqrt{\dfrac{\tau_a}{\pi}}\;C_1 $ and $K = \exp\left( - \int_0^{\tau} dt \left[\dfrac{(\dot{x}_t - g_t)^2}{4D} + \dfrac{1}{2} \dfrac{\partial g_t}{\partial x}  \right] \right)$.\\
We integrate the term with $\dot{\eta}^2$ using integration by parts
and we also directly integrate the term with $\dot{\eta}_t \eta_t$.
\begin{align*}
p(\underline{x}|x_0) = C_1' K \int \mathcal{D} \overline{\eta} \; \exp \left(\int_0^{\tau} dt \dfrac{\sqrt{2D_a}}{2D} \eta_t (\dot{x}_t - g_t) - \dfrac{1}{2} \int_0^{\tau} dt \int_0^{\tau} dt' \; \eta_t \hat{V}_{\tau}(t,t') \eta_{t'}  \right),
\end{align*}
where $\hat{V}$ is a differential operator defined as:
\begin{align*}
\hat{V}_{\tau}(t,t') = \delta(t-t') \left[-\tau_a^2 \partial_t^2 + (1+D_a/D) + \delta(t) (\tau_a - \tau_a^2 \partial_t) + \delta(t-\tau)(\tau_a + \tau_a^2 \partial_t) \right].
\end{align*}

Then, the path integral over the active noise paths $\eta(t)$ can be done exactly~\cite{Irreversibility_in_active_matter}. The idea is to first complete the square so that we can write the integral as:
\begin{align*}
\int \mathcal{D} \overline{\eta} \exp \left( \dfrac{1}{2} \int_0^{\tau} dt \int_0^{\tau} dt' [w_t^T \Gamma_{\tau}(t,t') w_{t'} - (\eta_t + \epsilon_t)^T V_{\tau}(t,t') (\eta_{t'} + \epsilon_{t'})] \right)
\end{align*}
where we don't know $\epsilon$ and $\Gamma_{\tau}$ yet, but they are used to complete the square, and $w_t$ is defined as $\dfrac{\sqrt{2 D_a}}{2D} (\dot{x}_t - g_t)$. 
The path integral over noises can be shifted to an integral over $\eta_t + \epsilon$ instead of $\eta$, and this change of variable has the identity as its Jacobian. Then, the
result we are looking for is:
\begin{small}
\begin{align}
p(\underline{x}|x_0)
& = C_1' K B \exp \left( \dfrac{1}{2} \int_0^{\tau} dt \int_0^{\tau} dt'\; w_t^T \Gamma_{\tau}(t,t') w_{t'} \right)
\label{ckb}
\end{align}
\end{small}
where we define $B$ as:
\begin{align}
B := \int \mathcal{D} \overline{\eta} \exp\left(-\frac{1}{2} \int_0^{\tau} dt \int_0^{\tau} dt' \; (\eta_t + \epsilon_t)^T V_{\tau}(t,t') (\eta_{t'} + \epsilon_{t'}) \right) 
\label{eq:B1}
\end{align}

\paragraph*{Calculation of $B$}
$\;$ \\
We still need to calculate the factor $B$ of Eq.\ref{eq:B1} so that we can get the complete normalization factor. 
Substituting $\hat{V}$ into the expression for $B$ we get:
\begin{align*}
B = & \int \mathcal{D}  \overline{\eta} \exp \bigg[ -\dfrac{1}{2} \int_0^{\tau} \int_0^{\tau} dt dt' \; \delta(t-t')  \big[ -\tau_a^2 \eta(t) \ddot{\eta}(t') + L \eta(t) \eta(t') + \\
& \;\;\; + \delta(t) ( \tau_a \eta(t)\eta(t') - \tau_a^2 \eta(t) \dot{\eta}(t')) + \delta(\tau) (\tau_a \eta(t) \eta(t') + \tau_a^2 \eta(t) \dot{\eta}(t')) \big] \bigg].
\end{align*}
Then, getting rid of the Dirac deltas by doing the integration results in:
\begin{align*}
\int \mathcal{D}  \overline{\eta} \; \exp \left[ \frac{1}{2} \int_0^{\tau} dt  \; \left( \tau_a^2 \eta(t) \ddot{\eta}(t) - L \eta(t)^2 \right) + \dfrac{1}{2} BT \right] 
\end{align*}
where $BT  = -\tau_a  \eta(0)^2  + \tau_a^2 \eta(0) \dot{\eta}(0) - \tau_a \eta(\tau)^2 - \tau_a^2 \eta(\tau) \dot{\eta}(\tau) $ are the boundary terms. 

Then, we can use integration by parts to solve $\int_0^{\tau} dt \; \eta(t) \ddot{\eta}(t) = \eta(\tau) \dot{\eta}(\tau) - \eta(0) \dot{\eta}(0) - \int_0^\tau dt\; \dot{\eta}(t)^2$. 
Therefore:
\begin{align*}
B &= \int \mathcal{D} \overline{\eta} \; \exp \left[ \dfrac{1}{2} \int_0^{\tau} dt  \; \left(-\tau_a^2 \dot{\eta}(t)^2 - L \eta(t)^2 \right) +  \dfrac{1}{2} \tau_a^2 \eta(\tau)\dot{\eta}(\tau) - \dfrac{1}{2} \tau_a^2 \eta(0) \dot{\eta}(0) + 
\dfrac{1}{2} BT   \right]  \\
& =  \int \mathcal{D}  \overline{\eta} \; \exp \left[ -\dfrac{1}{2} \int_0^{\tau} dt  \; \left(\tau_a^2 \dot{\eta}(t)^2 + L \eta(t)^2 \right) +  
\dfrac{1}{2} BT'   \right] 
\end{align*}
where $BT' =  BT +  \tau_a^2 \eta(\tau)\dot{\eta}(\tau) - \tau_a^2 \eta(0) \dot{\eta}(0) =   -\tau_a \eta(0)^2 - \tau_a \eta(\tau)^2$. We rewrite the term in the integral using 
$\tau_a^2 \dot{\eta}(t)^2 + L \eta(t)^2 =  (\tau_a \dot{\eta}(t) + \sqrt{L} \eta(t))^2 - 2\tau_a \sqrt{L}  \dot{\eta}(t) \eta(t).$
\begin{align*}
B = & \int \mathcal{D}  \overline{\eta} \; \exp \left[ - \dfrac{1}{2} \int_0^{\tau} dt  \; \left(   (\tau_a \dot{\eta}(t) + \sqrt{L} \eta(t))^2 - 2\tau_a \sqrt{L}  \dot{\eta}(t) \eta(t) \right) + \dfrac{1}{2} BT'  \right]  \\
 = & \int \mathcal{D}  \overline{\eta} \; \exp \left[ - \dfrac{1}{2} \int_0^{\tau} dt  \; \left(   \tau_a \dot{\eta}(t) + \sqrt{L} \eta(t) \right)^2 + \dfrac{1}{2} \tau_a \sqrt{L} \eta(\tau)^2 - \dfrac{1}{2} \tau_a \sqrt{L} \eta(0)^2
 + \dfrac{1}{2} BT'  \right].
\end{align*}
Therefore, 
\begin{align}
B = \int \mathcal{D}  \overline{\eta} \exp \left[ - \dfrac{1}{2} \int_0^{\tau} dt\; (\tau_a \dot{\eta}(t) + \sqrt{L} \eta(t))^2 + \dfrac{1}{2} BT'' \right],
\label{eq:casi}
\end{align}
where now $BT'' = BT' + \tau_a \sqrt{L} \eta(\tau)^2 - \tau_a \sqrt{L} \eta(0)^2 =  -\tau_a k_+ \eta(0)^2 - \tau_a k_- \eta(\tau)^2 $, with $k_{\pm} = 1 \pm \sqrt{L}$.

At this point, we can identify the integral in Eq.~\eqref{eq:casi}  as having the shape of an Onsager-Machlup integral for a system defined by the following Langevin equation
\begin{align}
\dot{y}(t) = - \dfrac{\sqrt{L}}{\tau_a}
y(t) + \dfrac{1}{ \tau_a} \xi(t).
\label{y}
\end{align}
For such a system, the probability of a given trajectory $\underline{y}$ is given by the Onsager-Machlup integral according to Eq.~\eqref{eq:machlup}:
\begin{align*}
p(\underline{y} | y_0)  = C_0'\exp \left(  -\dfrac{1}{2} \int_0^{\tau} dt \; (\tau_a \dot{y}(t) + \sqrt{L} y(t) )^2 \; + \dfrac{\sqrt{L} \tau }{2 \tau_a}  \right), 
\end{align*}
where $C_0' = \left( \dfrac{\tau_a}{  \sqrt{ 2 \pi \Delta t}} \right)^N$ is the normalization constant. Noting that this is very similar to the result we had for $B$ in Eq.~\eqref{eq:casi} is very useful, since we
know that $p(\underline{y}|y_0)$ is normalized, so that $\int \mathcal{D}\underline{y} \;p(\underline{y}|y_0) = 1$. However, we are still missing something before being able to use this, since the integral for $B$ is over $\overline{\eta}$, while the one on $y$ is over $\underline{y}$, so we need to include the initial point $y_0$ in the $y$ integral to make it comparable to the $\eta$ integral. 

From the Langevin equation for $y$, we know that the distribution of $y_0$ in steady state is a Gaussian with mean $0$ and standard deviation $1/(\sqrt{2 \tau_a} L^{1/4})$. Therefore, since the path integral is normalized, we have that:
\begin{align}
\sqrt{\dfrac{\tau_a}{\pi}} L^{1/4} \int \mathcal{D} \overline{y} \; p(\underline{y}|y_0)  e^{- \tau_a \sqrt{L} y(0)^2} = 1.
\label{eq:normalized}
\end{align}
We notice that this result is the same as we have for the $\eta$ process in Eq.~\eqref{eq:casi}, but with a term $\dfrac{\sqrt{L} \tau}{2 \tau_a} -\tau_a  \sqrt{L} \eta(0)^2$, and without the $\dfrac{1}{2} BT''$ term. We know that when doing a path integral of this quantity, the result is $1$ because it is normalized, therefore, we rewrite the $\eta$ integral of Eq.~\eqref{eq:casi} as:
\begin{small}
\begin{align}
\label{note}
\nonumber B  &=  \int \mathcal{D}  \overline{\eta} \exp \left[ - \dfrac{1}{2} \int_0^{\tau} dt\; (\tau_a \dot{\eta}(t) + \sqrt{L} \eta(t))^2 + \dfrac{1}{2} BT'' \right] \\
& = \int \mathcal{D}  \overline{\eta} \exp \left[ - \dfrac{1}{2} \int_0^\tau \; dt\; (\tau_a \dot{\eta}(t) + \sqrt{L} \eta(t))^2 \; + \dfrac{\sqrt{L}\tau}{2\tau_a} - \tau_a \sqrt{L} \eta(0)^2 \right] \exp \left[ \dfrac{1}{2} BT'' - \dfrac{\sqrt{L} \tau}{2\tau_a}  + \tau_a \sqrt{L} \eta(0)^2\right]
\end{align}
\end{small}

The integral of the first part is the one we have for the process $y$, so we know that it is normalized, and considering the $\sqrt{\dfrac{\tau_a}{\pi}} L^{1/4}$ factor in Eq.~\eqref{eq:normalized}, the result of this integral is $\dfrac{1}{C_0'} \sqrt{\dfrac{\pi}{\tau_a}} L^{-1/4}$. Therefore, we have:
\begin{align}
B &=  \sqrt{\dfrac{\pi}{\tau_a}} L^{-1/4} \; \dfrac{\langle e^{BT''/2\; +\; \tau_a \sqrt{L} \eta(0)^2} \rangle e^{-\tau \sqrt{L}/(2\tau_a)}}{C_0'}.
\label{eq:final_B}
\end{align}

\paragraph*{Putting the Results Together}
$\;$ \\ 

Now that we have the result for $B$, we can put everything together into equation \ref{ckb} to get $p(\underline{x}|x_0)$:
\begin{align*}
p(\underline{x}|x_0) = C_1' B \exp \left( - \int_0^{\tau} dt \left[  \dfrac{(\dot{x}_t - g_t)^2}{4D} + \dfrac{1}{2} \dfrac{\partial g_t}{\partial x}\right] +  \dfrac{D_a}{4D^2} \int_0^{\tau} dt \int_0^{\tau} dt' (\dot{x}_t - g_t) \Gamma_{\tau}(t,t') (\dot{x}_{t'}-g_{t'}) \right)
\end{align*}
with $C_1' B$ given by:
\begin{align*}
C_1' B & = \sqrt{\dfrac{\tau_a}{\pi}} L^{-1/4}  \left( \dfrac{1}{\sqrt{2\pi \Delta t}} \right)^{2N} \left( \dfrac{\tau_a}{\sqrt{2D}} \right)^N e^{\tau/(2\tau_a)} \sqrt{\dfrac{\pi}{\tau_a}}  \dfrac{\langle e^{BT''/2 + \tau_a \sqrt{L} \eta(0)^2} \rangle e^{-\tau \sqrt{L} / (2\tau_a)} }{\left( \dfrac{\tau_a}{\sqrt{2 \pi \Delta t}} \right)^N  } \\
& = L^{-1/4} \left( \dfrac{1}{\sqrt{4 \pi D \Delta t} } \right)^{N} e^{\tau k_- /(2\tau_a)} \langle e^{BT''/2 + \tau_a \sqrt{L} \eta(0)^2} \rangle. \\
\end{align*}
To finish the calculation, we still need to obtain the expected value $\langle e^{BT''/2 + \tau_a \sqrt{L} \eta(0)^2 } \rangle$, where $BT'' = -\tau_a k_+ \eta(0)^2 - \tau_a k_- \eta(\tau)^2$, with $k_{\pm} = 1 \pm \sqrt{L}$.
This is:
\begin{align*}
\langle e^{-\tau_a k_- \eta_0^2 / 2 - \tau_a k_- \eta_{\tau}^2 /2} \rangle = \langle e^{q (\eta_0^2 + \eta_{\tau}^2)} \rangle,
\end{align*}
with $q:= -\tau_a k_-/2$ (the minus sign is so that $q$ is positive, because $k_-$ is always negative).  Notice that, as we said earlier, the expected value is done by interpreting the variable $\eta$ to follow the equation~\ref{y} and not the equation we had originally for $\eta$. Therefore, the statistics of $\eta$ needs to be obtained from equation~\ref{y} and the result is:
\begin{itemize}
    \item $\sigma_0^2 := \langle \eta_0^2 \rangle =  \dfrac{1}{2 \sqrt{L} \tau_a}$ \\
    \item $\sigma_{\tau}^2 := \langle \eta_{\tau}^2 \rangle =  \dfrac{1}{2 \sqrt{L} \tau_a}$ \\
    \item  $\rho :=  \dfrac{\langle \eta_0 \eta_{\tau} \rangle }{\sigma_0 \sigma_{\tau}}  = e^{- \sqrt{L}\tau/\tau_a}$
\end{itemize}

Then, using the bivariate Gaussian distribution, we have that: 
\begin{align*}
\langle e^{q (\eta_0^2 + \eta_{\tau}^2)} \rangle = \dfrac{1}{2 \pi \sigma_0 \sigma_{\tau} \sqrt{1-\rho^2}}  \int_{-\infty}^{\infty} \int_{-\infty}^{\infty}  \exp \left[ -\dfrac{1}{2(1-\rho^2)} \left( \dfrac{\eta_0^2}{\sigma_0^2} + \dfrac{\eta_{\tau}^2}{\sigma_{\tau}^2} - 2 \rho \dfrac{\eta_0 \eta_{\tau}}{\sigma_0 \sigma_{\tau}}  \right)  \right] e^{q \eta_0^2 + q\eta_{\tau}^2} \; d \eta_0 d \eta_{\tau},
\end{align*}
Therefore:
\begin{align*}
\langle e^{q (\eta_0^2 + \eta_{\tau}^2)} \rangle &= \dfrac{\sqrt{L} \tau_a}{ \pi \sqrt{1-\rho^2}}  \int_{-\infty}^{\infty} \int_{-\infty}^{\infty}  \exp \left[ -\dfrac{1}{2(1-\rho^2)} \left( 2 \sqrt{L} \tau_a \eta_0^2 + 2 \sqrt{L}  \tau_a \eta_{\tau}^2 - 4  \sqrt{L}  \tau_a \rho \;\eta_0 \eta_{\tau}  \right) + q \eta_0^2 + q \eta_{\tau}^2 \right]  \; d \eta_0 d \eta_{\tau} \\
& = \dfrac{\sqrt{L} \tau_a}{ \pi \sqrt{1-\rho^2}}  \int_{-\infty}^{\infty} \int_{-\infty}^{\infty}  \exp \bigg\{ -\dfrac{1}{2} \begin{pmatrix} 
\eta_0 & \eta_{\tau}
\end{pmatrix}
\begin{pmatrix}
\dfrac{2 \sqrt{L} \tau_a}{1-\rho^2} - 2q & - \dfrac{2 \sqrt{L}  \tau_a \rho}{1-\rho^2} \\
 - \dfrac{2 \sqrt{L}  \tau_a \rho}{1-\rho^2} & \dfrac{2 \sqrt{L}  \tau_a}{1-\rho^2} - 2q
\end{pmatrix} \begin{pmatrix}
\eta_0 \\ \eta_{\tau}
\end{pmatrix}
\bigg\}\; d \eta_0 d \eta_{\tau}
\end{align*}
The result of this Gaussian integral is $\dfrac{2\pi }{\sqrt{\det A}}$, where $A$ is the matrix inside the exponential. The determinant we have is:
\begin{align*}
\det A  = \dfrac{4 \sqrt{L} \tau_a (\sqrt{L} \tau_a - 2  q) + 4q^2 (1-\rho^2)  }{1-\rho^2}
\end{align*}
where again $q = -\tau_a k_- /2$, $k_- = 1- \sqrt{L} = 1-\sqrt{1+D_a/D}$, and  $\rho = e^{-\tau / \tau_a}$. Then, we can substitute $q$ and $k_-$ to get:
\begin{align*}
\det A  = \dfrac{4 \sqrt{L} + (1-\sqrt{L})^2 (1-\rho^2) }{1-\rho^2}  \tau_a^2 
\end{align*}
Therefore, the expected value is:
\begin{align*}
\langle e^{q(\eta_0^2 + \eta_{\tau}^2)} \rangle & =  \dfrac{\sqrt{L} \tau_a}{\pi \sqrt{1-\rho^2}}
\dfrac{2\pi }{\sqrt{\det A}} \\
& = \dfrac{2\sqrt{L}}{\sqrt{ 4 \sqrt{L} + (1-\sqrt{L})^2 (1-\rho^2)  }  }\\
\end{align*}
Finally, the constant $C_1' B$ in front of the exponential in the expression for $p(\underline{x}|x_0)$ is
\begin{align*}
C_1' B = L^{1/4} \left( \dfrac{1}{\sqrt{4 \pi D \Delta t}} \right)^N e^{\tau k_- / (2\tau_a)} \dfrac{2}{\sqrt{ 4 \sqrt{L} + (1-\sqrt{L})^2 (1-\rho^2)  }  }
\end{align*}
with $L = 1+D_a/D$, $k_- = 1-\sqrt{L}$, and $\rho = e^{-\sqrt{L} \tau/\tau_a}$.

\paragraph*{Final Result}
$\;$ \\ \\
The final result is:
\begin{align*}
p(\underline{x}|x_0) = C_1' B \exp \left( - \int_0^{\tau} dt \left[  \dfrac{(\dot{x}_t - g_t)^2}{4D} + \dfrac{1}{2} \dfrac{\partial g_t}{\partial x}\right] +  \dfrac{D_a}{4D^2} \int_0^{\tau} dt \int_0^{\tau} dt' (\dot{x}_t - g_t) \Gamma_{\tau}(t,t') (\dot{x}_{t'}-g_{t'}) \right),
\end{align*}
with $C_1' B$ given by:
\begin{align*}
C_1' B = L^{1/4} \left( \dfrac{1}{\sqrt{4 \pi D \Delta t}} \right)^N e^{\tau k_- / (2\tau_a)} \dfrac{2}{\sqrt{ 4 \sqrt{L} + (1-\sqrt{L})^2 (1-\rho^2)  }  } ,
\end{align*}
and  $L = 1+D_a/D$, $k_{\pm} = 1\pm\sqrt{L}$, $\rho = e^{-\sqrt{L} \tau/\tau_a}$. \\

We are just missing to mention the value of $\Gamma_{\tau}$ which was introduced before Eq.~\eqref{ckb} for completing the square in the integral. We don't show it here, but the result is given in~\cite{active_brownian} and it is:
\begin{align}
\Gamma_{\tau}(t,t') = \left( \dfrac{1}{2 \tau_a \sqrt{L}} \right) \dfrac{k_+^2 e^{-\sqrt{L} |t-t'|/\tau_a} + k_-^2 e^{-\sqrt{L}(2 \tau - |t-t'|)/\tau_a} - k_+k_- \left[ e^{-\sqrt{L} (t+t')/\tau_a} + e^{-\sqrt{L}(2\tau - t-t')/\tau_a} \right]}{k_+^2 - k_-^2 e^{-2 \sqrt{L}\tau /\tau_a}}.
\label{eq: gamma}
\end{align}

Then, we can substitute $P(\underline{x}|D_a = 0)$ and $P(\underline{x}|D_a  = D_a')$ into
Eq.~\eqref{eq:bayes_result}
to calculate $P(D_a = D_a' | \underline{x})$ for any trajectory. When doing so, some things cancel out and we have our final result:

\begin{align}
P(D_a = D_a' | \underline{x} ) = \dfrac{1}{1+   C \exp \bigg\{ - \dfrac{D_a}{4D^2} \displaystyle \int_0^{\tau} dt \int_0^{\tau}\;  dt' (\dot{x}_t - g_t) \Gamma_{\tau}(t,t') (\dot{x}_{t'}-g_{t'})  \;  \bigg\} },
\label{eq:final_theo2}
\end{align}
with $C$ given by
\begin{align*}
C = \dfrac{1}{2} L^{-1/4} e^{- \tau k_- / (2\tau_a)}  \sqrt{ 4 \sqrt{L} + (1-\sqrt{L})^2 (1-\rho^2)  }.
\end{align*}

\section{Experimental setup}
\subsection*{Cell Culture}
MCF-7 cells were used for the experiments and were a gift from Professor Jesper Nylandsted, Danish Cancer Institute. MCF-7 cells were grown in vented T25 flasks (BD Falcon) in a sterile environment at 37 °C in a humidified $5\%$ CO$_2$ incubator. The cells were passaged at around $80\%$ confluence. Cells were grown in DMEM ([+] $4.5 g/L$ D-Glucose, L-Glutamine, [+] $110 mg/L$ Sodium Pyruvate (Gibco)) with $20 \mu M$ HEPES. Onde day prior to experiments, cells were seeded at a concentration of $700,000$ cells / $ml$ on micorscope cover glasses ($24 \times 60 mm$, thickness $0.13$ to $0.16 mm$) containing ibidi culture inserts (2-well)). 1h before the experiment, the ibidi insert was removed and the samples were kept in the incubator such that the cells were allowed to migrate into the cell free void. Prior to imaging, a second cover glass ($24 \times 50 mm$) was mounted on top using double-sided tape along the edges to form a sealed chamber. This chamber was then placed on the microscope stage, positioned between a water immersion objective and an oil immersion condenser.

\subsection*{Optical Tweezers}
Experiments were conducted using a LUMICKS C-Trap optical tweezers system. A near-infrared trapping laser was focused through a 60× water immersion confocal objective (Nikon). A high-resolution piezo stage enabled for precise positioning of the optical trap directly above the target granule, allowing stable trapping within the viscoelastic cytoplasmic environment. The forward scattering light from the trapping laser was collected by an oil immersion condenser and directed to a position sensitive detector (PSD), which recorded the position of the granule at a sampling rate of 70 kHz.
Each granule was tracked for approximately 3 to 5 seconds to optimize the signal-to-noise ratio. No physiological damage was observed during the measurement period, indicating that the trapping laser did not induce detectable phototoxic or heating effects under the experimental conditions. All measurements were performed at room temperature (approximately 37\,\textdegree{}C). 


\newpage
\end{widetext}

\end{document}